\documentclass[12pt]{article}
\usepackage{amssymb}
\usepackage{amsmath}
\usepackage{amstext,amsthm,amscd,mathrsfs}
\usepackage[top=60pt,bottom=70pt,left=75pt,right=75pt]{geometry}
\usepackage{setspace}
\usepackage{sectsty}

\newcommand\g{\gamma}
\newcommand\de{\delta}
\newcommand\e{\epsilon}
\newcommand\G{\Gamma}
\newcommand\D{\Delta}

\newcommand\Hv{H^*}
\newcommand{\script}{\scriptscriptstyle}
\newcommand{\w}{\mathrm{w}}
\newcommand{\ii}{\mathrm{i}}
\newcommand{\tr}{\mathrm{tr}}
\newcommand{\Real}{\mathrm{Re}}
\newcommand{\Imag}{\mathrm{Im}}
\providecommand{\norm}[1]{\lVert#1\rVert}

\sectionfont{\large}
\subsectionfont{\normalsize \bfseries}

\begin{document}

\begin{titlepage}

\begin{center}

\today

\vskip 2 cm 

\vskip .5 cm 
{ \large \bfseries A note on supersymmetric AdS$_6$ solutions  \\[3mm]  of massive type IIA supergravity }

\vskip 1.8 cm
{ Achilleas Passias }

\vskip 1cm
\textit{ \small Department of Mathematics, King's College, London, \\ The Strand, London WC2R 2LS,  United Kingdom \\ achilleas.passias@kcl.ac.uk}

\end{center}

\vskip 3 cm

\begin{abstract}

\noindent
Motivated by the AdS$_6$/CFT$_5$ correspondence, 
we study general supersymmetric solutions of massive type IIA supergravity, consisting of a warped product of six-dimensional anti-de Sitter space 
AdS$_6$ with a four-dimensional Riemannian manifold $M_4$, and fluxes compatible with the $SO(5,2)$ symmetry. We find that the only local solution of this 
form is the warped AdS$_6 \times_\w  S^4$ solution, originally discovered by Brandhuber and Oz. We discuss the supersymmetric properties of this solution.

\end{abstract}

\end{titlepage}

\pagestyle{plain}
\setcounter{page}{1}
\newcounter{bean}
\onehalfspacing

%\tableofcontents

\section{Introduction}

The AdS/CFT correspondence \cite{Maldacena} motivates the study of supersymmetric backgrounds of string theory, whose geometry possesses an $SO(p-1,2)$ symmetry. 
There is an extensive literature  on AdS$_p$  solutions of (massive) type IIA, type IIB  and eleven-dimensional supergravity for $p=2$  to  $p=5$ 
(for comprehensive studies see 
\cite{Martelli:2003ki,AdS5_M,Lin:2004nb,Tsimpis,AdS5_IIB,AdS4_M}), while for $p =7$ the only known supersymmetric background is the AdS$_7 \times S^4$ Freund-Rubin  solution of eleven-dimensional supergravity. It is therefore natural  
to investigate the existence of supersymmetric AdS$_6$ solutions, which are expected to be dual to superconformal fixed points in five dimensions \cite{Seiberg:1996bd,Morrison:1996xf}. Recent studies of five-dimensional superconformal
 field theories and the AdS$_6$/CFT$_5$ correspondence \cite{Rodriguez-Gomez, KKL, Jafferis} further motivate the search of such solutions.

In reference \cite{Oz} a (singular) warped AdS$_6 \times_\w S^4$ solution was found in massive type IIA supergravity \cite{Romans}, arising
 as the near horizon limit of a localised D4-D8 brane configuration. Prior to \cite{Oz}, the existence of this solution was anticipated in \cite{Ferrara}, 
where the connection to Romans ${F(4)}$ gauged supergravity \cite{RomansF(4)} in six dimensions and its boundary superconformal singleton theory was explored. 
The AdS$_6 \times_\w S^4$ solution was also recovered in \cite{Pope}, where the ${F(4)}$ gauged supergravity was obtained upon Kaluza-Klein reduction of the bosonic 
sector of massive IIA supergravity on $S^4$. The $\mathrm{SCFT_5}$ dual to 
this AdS$_6 \times_\w S^4$  gravitational background was further studied in \cite{D'Auria}. A more detailed description of the  AdS$_6$/CFT$_5$ correspondence has 
recently emerged with the work of \cite{Rodriguez-Gomez, Jafferis}.

The superconformal group in five dimensions is ${F(4)}$ and its bosonic subgroup is ${SO(5,2)}$ $\times$ ${SU(2)_R}$ \cite{Nahm}. As discussed in \cite{Oz}, the warped nature of 
the AdS$_6 \times_\w S^4$ metric reduces the ${SO(5)}$ isometry group of $S^4$ to ${SO(4)} \simeq {SU(2)_R \times SU(2)}$ where ${SU(2)_R}$ is the R-symmetry group of $\mathrm{F(4)}$, realised as a subgroup of this isometry. 

In the present work we study the most general supersymmetric AdS$_6 \times_\w M_4$ background of massive type IIA supergravity. We analyze the constraints imposed by supersymmetry on the geometry and the fluxes and conclude that the only background of this form is the 
known AdS$_6 \times_\w S^4$. In the course of the analysis, we discuss how the ${SU(2)_R}$ subgroup of the ${SO(5,2) \times SO(4)}$ isometry group of this solution 
is realised in terms of Killing vectors constructed as bilinears of the
 Killing spinors, as expected from general arguments \cite{FigueroaO'Farrill:1999va,FigueroaO'Farrill:2007ar}. Furthermore,
 we construct the Killing spinors of the solution. The solution preserves 16 supersymmetries in accordance with the number of supercharges of the ${F(4)}$ superalgebra.

The rest of the note is organised as follows. In section \ref{massiveIIA} we summarise some elements of the massive type IIA supergravity theory, mainly to establish our notation and conventions. In section \ref{susyconditions} we derive the reduced conditions for supersymmetry on $M_4$. In section \ref{analysis} we present the analysis of the supersymmetry conditions and in section \ref{solution} we reproduce 
the AdS$_6 \times_\w S^4$ solution and study its supersymmetric properties. The appendices include some technical material used in the main text.

\section{Massive type IIA supergravity}\label{massiveIIA}

In this section we briefly review some  elements of massive type IIA supergravity that are necessary for our analysis, 
following the conventions of reference \cite{Tsimpis}.

The bosonic sector of massive type IIA supergravity consists of the metric $g^{(10)}$, the dilaton $\phi$, 
a massive 2-form field $B'$ and a 3-form field $C'$. The field strengths of the form fields are defined as
\begin{equation}
H =d B' \, , \qquad
dG =d C' + m B'\wedge B'
\end{equation}
and the corresponding  Bianchi identities are
\begin{subequations}
\begin{align}
&d H = 0 \\
&d G = 2mB' \wedge H ~ .
\end{align}
\end{subequations}
The massless limit $m \rightarrow 0$ is retrieved after introducing the field redefinitions
\begin{subequations}
\begin{align}\label{B'}
mB'  &= mB + \tfrac{1}{2} dA \\  
mC' &= mC - \tfrac{1}{4} A \wedge dA \, ,
\end{align}
\end{subequations}
where $A$, $C$ and $B$ become the RR 1-form, the RR 3-form and the NS 2-form potentials of type IIA supergravity. 

The equations of motion of the theory are
\begin{subequations}
\begin{align}
& 0 = R_{\script{MN}} - \frac{1}{2} \nabla_{\script{M}} \phi \nabla_{\script{N}} \phi 
  -\frac{1}{12} e^{\phi/2} G_{\script{MPQR}}G_{\script{N}}{}^{\script{PQR}}
  +\frac{1}{128}e^{\phi/2} g^{(10)}_{\script{MN}}G^2
  -\frac{1}{4} e^{-\phi} H_{\script{MPQ}}H_{\script{N}}{}^{\script{PQ}}              \nonumber                                             \\
& ~~~
  +\frac{1}{48}e^{-\phi}g^{(10)}_{\script{MN}}H^2 
  -2m^2e^{3\phi/2} B'_{\script{MP}}B'_{\script{N}}{}^{\script{P}} + \frac{m^2}{8}e^{3\phi/2}g^{(10)}_{\script{MN}}(B')^2
  -\frac{m^2}{4}e^{5\phi/2}g^{(10)}_{\script{MN}}                                                                                                                          \\
& 0 =\nabla^2\phi-\frac{1}{96} e^{\phi/2}G^2 + \frac{1}{12} e^{-\phi}H^2
   - \frac{3m^2}{2}e^{3\phi/2}(B')^2 - 5m^2e^{5\phi/2}                                                                                                     \\     
& 0 = d(e^{-\phi}*H) -\frac{1}{2} G \wedge G + 2m e^{\phi/2} B' \wedge *G + 4m^2 e^{3\phi/2} *B'                             \\
& 0 = d(e^{\phi/2}*G) - H \wedge G ~ .
\end{align}
\end{subequations}
For a  bosonic background of a supergravity theory to be supersymmetric, the variation of the fermionic fields under a supersymmetry transformation must vanish i.e.\ there exists a Killing spinor $\e$ such that $\de_{\e}\text{(fermionic field)} = 0$. The fermionic fields of massive type IIA supergravity are the gravitino $\Psi_ {\scriptscriptstyle{M}}$ and the dilatino $\lambda$. The variation of the gravitino under a supersymmetry transformation is $\delta \Psi_ {\scriptscriptstyle{M}} = \tilde{\nabla}_{\scriptscriptstyle{M}} \e$ , where $\tilde{\nabla}_{\scriptscriptstyle{M}}$ is the supercovariant derivative
\begin{equation}
\begin{split}
\tilde{\nabla}_{\script{M}} &=
\nabla_{\script{M}} 
 - \frac{m e^{5\phi/4}}{16} \G_{\script{M}} 
 - \frac{m e^{3\phi/4}}{32} B'_{\script{NP}} 
   \left( \G_{\script{M}}{}^{\script{NP}} - 14\, \de_{\script{M}}{}^{\script{N}} \G^{\script{P}} \right) \G_{11} \\
&+ \frac{e^{-\phi/2}}{96}H_{\script{NPQ}} 
   \left( \G_{\script{M}}{}^{\script{NPQ}} - 9 \, \de_{\script{M}}{}^{\script{N}} \G^{\script{PQ}} \right)\G_{11} 
 + \frac{e^{\phi/4}}{256}G_{\script{NPQR}} 
   \left( \G_{\script{M}}{}^{\script{NPQR}} - \tfrac{20}{3} \, \de_{\script{M}}{}^{\script{N}}\G^{\script{PQR}} \right) \, ,
\end{split}
\end{equation}
while the dilatino variation reads
\begin{equation}
\begin{split}
\de\lambda =  [ \,  - \frac{1}{2} \partial_{\script{M}}  \phi \, \G^{\script{M}} &- \frac{5m e^{5\phi/4}}{4}
                                + \frac{3me^{3\phi/4}}{8} B'_{\script{MN}} \G^{\script{MN}} \G_{11} \\
                       &       + \frac{e^{-\phi/2}}{24} H_{\script{MNP}} \G^{\script{MNP}} \G_{11}
                                 - \frac{e^{\phi/4}}{192} G_{\script{MNPQ}} \G^{\script{MNPQ}} \, ] \, \e ~ .
\end{split}
\end{equation}
The supersymmetry parameter $\e$ is a Majorana spinor of $\mathrm{Spin(9,1)}$.

As shown in \cite{Tsimpis}, upon imposing  the Bianchi identities and the equations of motion of the fluxes, supersymmetry implies that the dilaton and Einstein equations are satisfied, provided $E_{0\scriptscriptstyle{M}} = 0$ for $M \neq 0$, where $E_{\script{MN}} = 0$ are the Einstein equations.

\section{AdS$_6 \times_\w M_4$ backgrounds}

\label{susyconditions}

We consider the most general supersymmetric bosonic  background of massive type IIA supergravity that is invariant under the action of  ${SO(5,2)}$. Accordingly, the ten-dimensional metric (in the Einstein frame) is assumed to be of the form of a warped sum of a metric 
$g_{\mathrm{AdS}_6}$ on AdS$_6$ and an arbitrary four-dimensional  Riemannian metric $g$ as
\begin{equation}
g^{(10)} =  e^{2\Delta} \left( \,g_{\mathrm{AdS}_6} + g \, \right) ~ . 
\end{equation} 
The warp factor $\Delta$ is a function on $M_4$. 
The signature of the ten-dimensional metric is $(-,+, \dots, +)$. In conformance with the $SO(5,2)$ symmetry, the fluxes have non-vanishing components only on $M_4$. Furthermore, the equation of motion for $G$ sets 
\begin{equation}
G = \mu  e^{-\phi/2 - 2\D} \mathrm{vol}_4 \, ,
\end{equation}
where $\mu$ is a constant and $\mathrm{vol}_4$ the Riemannian volume form. It will prove convenient to introduce an abbreviation for the dual of $H$ in four dimensions: $\Hv \equiv *_4 H$.
Then, the equation of motion for $H$ becomes 
\begin{equation}\label{eom}
d(e^{-\phi + 4\D} \Hv) + 2 \mu m B' + 4m^2 e^{3\phi/2 + 6\D} *_4B' = 0 ~ .
\end{equation}

In order to study the conditions imposed by supersymmetry on  the fluxes and the geometry of $M_4$, we decompose the supersymmetry parameter $\e$  in terms of $\mathrm{Spin(5,1)}$ and  $\mathrm{Spin(4)}$ spinors. The decomposition ansatz for the $\mathrm{Spin(9,1)}$ Majorana spinor $\e$ is
\begin{equation}\label{spinoransatz}
\epsilon  = \sum_{i=1}^2 \psi_i^+ \otimes e^{\D/2} \eta_i + \sum_{i=1}^2 \psi_i ^-\otimes e^{\D/2} \xi_i  ~ .
\end{equation}
Here  $\psi^+_i$, $\psi^-_i$ are symplectic-Majorana Weyl spinors of $\mathrm{Spin(5,1)}$, (a plus or minus index indicates the chirality  of the spinor) whereas  $\eta_i$ and  $\xi_i$ are symplectic-Majorana Dirac spinors of $\mathrm{Spin(4)}$. The summation over the symplectic-Majorana indices $i$ ensures that $\e$ satisfies a Majorana condition. The factor $e^{\D/2}$ is included for later convenience. The $\psi^+_i$ and $\psi^-_i$
 spinors on AdS$_6$ satisfy the Killing spinor equations
\footnote{
In principle one can consider the more general Killing spinor equations   $\nabla_\mu \psi_i^\pm =  \Lambda \g_\mu  \sum_j W_{ij} \psi_j^\mp $ but by a re-definition of 
$\psi^-_i$,  $W$ can be set equal to the identity.}
\begin{equation}
\nabla_\mu \psi_i^\pm = \Lambda  \g_\mu \psi_i^\mp ~ .
\end{equation}
The Ricci curvature of the AdS$_6$ space thus defined is $\mathrm{Ric} = - 5\cdot (2\Lambda)^2 g_{\mathrm{AdS}_6}$ . The spinor decomposition \eqref{spinoransatz} implies that for $\Lambda \neq 0$ both $\eta_i$ and $\xi_i$ spinors have to be nonzero. Furthermore, the chiral spinors  $\eta^+_i$  and  $\eta^-_i$  form a basis for the representations of positive and negative chirality of $\mathrm{Spin(4)}$ and thus $\xi_i$ can be expanded in terms of $\eta_i$ as 
\begin{equation}\label{expansion}
\begin{pmatrix} \xi^+_1 \\ \xi^+_2 \end{pmatrix} 
= \begin{pmatrix} a_1 & a_2 \\ -a_2^* & a_1^* \end{pmatrix} \begin{pmatrix} \eta^+_1 \\ \eta^+_2 \end{pmatrix}
\, , \qquad
\begin{pmatrix} \xi^-_1 \\ \xi^-_2 \end{pmatrix} 
= \begin{pmatrix} b_1 & b_2 \\ -b_2^* & b_1^* \end{pmatrix} \begin{pmatrix} \eta^-_1 \\ \eta^-_2 \end{pmatrix}
\end{equation}
where $a_1,a_2,b_1,b_2$ are complex functions. A review of spinors in $4$ and $5+1$ dimensions is included in appendix \ref{cliffspin}.

Decomposing the Clifford algebra $\mathrm{Cliff(9,1)}$ as $\mathrm{Cliff(5,1)} \otimes \mathrm{Cliff(4,0)}$ (c.f.\ appendix \ref{cliffspin}) and substituting the ans\"{a}tze in the gravitino and dilatino supersymmetry variations, yields a set of conditions on the $\mathrm{Spin(4)}$ spinors. We obtain
four algebraic conditions
\begin{subequations}\label{algebraic}
\begin{align}
[( \partial_m \phi - 12 \partial_m \D + 2 e^{\Phi_3} \Hv_m ) \g^m 
+ 4 m e^{\Phi_1} - 2 \mu e^{\Phi_4}\g_5 ] \eta_i  
- 24 \Lambda \gamma_5 \xi_i = 0 \label{alg1} \\
[( \partial_m \phi - 12 \partial_m \D - 2 e^{\Phi_3} \Hv_m ) \g^m 
 + 4 m e^{\Phi_1} - 2 \mu e^{\Phi_4}\g_5 ] \xi_i 
- 24 \Lambda \gamma_5 \eta_i = 0 \label{alg2} \\
[( 4 \partial_m \phi + 2 e^{\Phi_3} \Hv_m ) \g^m 
 + 10 m e^{\Phi_1}  - 3m e^{\Phi_2} B'_{mn} \g^{mn}\g_5 
 + \mu e^{\Phi_4} \g_5 ] \eta_i = 0  \label{alg3} \\ 
[ (4 \partial_m \phi - 2 e^{\Phi_3} \Hv_m ) \g^m
 + 10 m e^{\Phi_1}  + 3m e^{\Phi_2} B'_{mn} \g^{mn} \g_5 
 +\mu e^{\Phi_4} \g_5 ] \xi_i  = 0     \label{alg4}
\end{align}
\end{subequations}
from the dilatino variation and the AdS$_6$ components of the gravitino variation and two differential conditions
\begin{subequations}\label{differential}
\begin{align}
\nabla_m \eta_i   + \frac{m e^{\Phi_2}}{2} B'_{mn} \g^n \g_5 \eta_i  - \frac{\mu e^{\Phi_4} }{4}\g_m \g_5 \eta_i    
   - \frac{e^{\Phi_3}}{4} \Hv_n \g^n {}_m \eta_i  + \frac{e^{\Phi_3}}{8} \Hv_m \eta_i -  \Lambda \g_m \g_5 \xi_i = 0  \\
\nabla_m \xi_i  - \frac{m e^{\Phi_2}}{2} B'_{mn} \g^n \g_5 \xi_i - \frac{\mu e^{\Phi_4} }{4} \g_m \g_5 \xi_i
   + \frac{e^{\Phi_3}}{4} \Hv_n \g^n{}_m \xi_i - \frac{e^{\Phi_3}}{8} \Hv_m \xi_i  - \Lambda \g_m \g_5 \eta_i = 0  
\end{align}
\end{subequations}
from the $M_4$ components of the gravitino variation.
In the above equations we have introduced the exponents
\begin{equation*}
\Phi_1 \equiv \frac{5}{4}\phi + \D, \quad \Phi_2 \equiv \frac{3}{4}\phi - \D, \quad \Phi_3 \equiv -\frac{1}{2}\phi - 2\D, \quad  \Phi_4 \equiv -\frac{1}{4}\phi - 5\D ~ .
\end{equation*}

\section{Analysis of the supersymmetry conditions}

\label{analysis}

The strategy for analysing equations \eqref{algebraic} and \eqref{differential} is 
to translate them into algebraic and differential equations obeyed by bilinears 
of $\eta_i$ and $\xi_i$. We begin by introducing a set of (real) spinor bilinears which 
appear in our analysis. The symmetry and reality properties of bilinears of $\mathrm{Spin(4)}$ 
spinors are presented in appendix \ref{bilinearsproperties}.
\begin{center}
\begin{tabular}
{ |@{\hspace{.2cm}}l@{\hspace{.2cm}}
  |@{\hspace{.2cm}}l@{\hspace{.2cm}}
  l@{\hspace{.2cm}}
  |@{\hspace{.2cm}}l@{\hspace{.2cm}}| }
\hline
{\bf scalars}  & {\bf 1-forms} && {\bf 2-forms} 
\\ \hline \hline 
$\hat{s}_+ := \eta^\dagger_i \eta_i = \norm{\eta_i}^2 $    &  
$V^A_+ := \tfrac{1}{4} \tr \, \eta^\dagger \g_5  \g_{(1)} \ii \sigma^A \eta$  & 
$K^A := \tfrac{1}{2} \tr \, \eta^\dagger \g_5 \g_{(1)} \ii \sigma^A \xi$      &
$\mathrm{J}^A_+ := \tfrac{1}{2} \tr \,\eta^\dagger \g_{(2)} \ii \sigma^A \eta$ 
\\ [.1cm]
$\hat{s}_- := \xi^\dagger_i \xi_i = \norm{\xi_i}^2$               &
$V_+^4 := \tfrac{1}{4} \tr \, \eta^\dagger \g_{(1)} \eta$  &
$K := \tfrac{1}{2} \tr \, \eta^\dagger \g_5\g_{(1)} \xi$   &
$\mathrm{J}^A_- := \tfrac{1}{2} \tr \, \xi^\dagger \g_{(2)} \ii \sigma^A \xi$  
\\ [.1cm]
$s_+ := \eta^\dagger_i \g_5 \eta_i $                              &  
$V_-^4 := \tfrac{1}{4}\tr \, \xi^\dagger \g_{(1)} \xi$     &
$K' := \tfrac{1}{2} \tr \, \eta^\dagger \g_{(1)} \xi$ &
\\ [.1cm]
$s_- := \xi^\dagger_i \g_5 \xi_i $   &&& \\ [.1cm]
\hline                                                          
\end{tabular} 
\end{center}
In the above expressions we use the notation
\begin{equation}
\g_{(n)} = \frac{1}{n!} \g_{m_1 m_2 \dots m_n} dy^{m_1} \wedge dy^{m_2} \wedge \dots dy^{m_n}
\end{equation}
where $y^m$ are coordinates on $M_4$. The index $A \in \{1,2,3 \}$ and $\sigma^A$ are the Pauli matrices 
acting on the symplectic-Majorana indices of the spinors; the trace $\mathrm{tr}$ is also over the 
symplectic-Majorana indices, e.g.\ 
$V^3_+ = \frac{\ii}{4} (\eta_1^\dagger \g_5  \g_{(1)} \eta_1 - \eta_2^\dagger \g_5  \g_{(1)} \eta_2)$ . 
A plus subscript is used to denote bilinears of $\eta_i$ and a minus subscript bilinears of $\xi_i$.\footnote{Given a symplectic-Majorana Dirac spinor of $\mathrm{Spin(4)}$, $\hat{s}, s, V^A, V^4, \mathrm{J}^A$ are all the bilinears one can construct.}

Application of Fierz identities yields the following algebraic relations for the bilinears 
of a Dirac symplectic-Majorana spinor e.g. $\eta_i$
\begin{subequations}
\begin{align}
g^{mn} V_m^a V_n^b 
&= \tfrac{1}{4} (\hat{s}^2 - s^2) \, \delta^{ab} \label{orthogonal} \\
\tfrac{1}{4} (\hat{s}^2 - s^2) \mathrm{J}^A 
&= \hat{s} \, \tfrac{1}{2} \e^{ABC} V^B \wedge V^C + s \, V^A\wedge V^4 \, , \label{JV}
\end{align}
\end{subequations}
where $a,b \in \{1,2,3,4\}$. Repeated indices $A,B,C, \dots$ here and henceforth are summed over. 
The Levi-Civita symbol $\e^{ABC}$ is normalised as $\e^{123} = 1$. Equation \eqref{orthogonal} implies 
that $V^a$ can be used to define an orthonormal frame on $M_4$ 
i.e.\ $\eta_i$ defines an identity structure. This is expected since the chiral components of $\eta_i$ span 
the chiral representations of $\mathrm{Spin(4)}$ or equivalently the isotropy group of $\eta_i$ 
in  $\mathrm{Spin(4)}$ is the identity $\mathbb{I}$. 

We proceed by stating a key set of relations for the scalar bilinears, deduced from the supersymmetry conditions. 
From \eqref{alg3} and \eqref{alg4} we derive
\begin{equation}\label{key}
10 m \, e^{\Phi_1} s_\pm + \mu \, e^{\Phi_4} \hat{s}_\pm = 0 ~ .
\end{equation}
The above equations imply that $\mu = 0$ if and only if $s_\pm = 0$. 
In the following subsections the two cases $\mu \neq 0$ and $\mu = 0$ will be considered separately.
A judicious use of \eqref{algebraic} and \eqref{differential} yields the following differential equations for the scalars
\begin{subequations}\label{ders0}
\begin{align}
\mp d \hat{s}_\pm 
&= \frac{e^{\Phi_3} \hat{s}_\pm }{4} \Hv  + 2 \Lambda K \\
e^{\phi/8 - 3 \Delta/2} d ( e^{-\phi/8 + 3 \Delta/2} \hat{s}_\pm) 
&= m e^{\Phi_1} V_\pm^4 \pm \Lambda K 
\end{align}
\end{subequations}
and
\begin{equation}\label{ders1}
e^{-3\phi/8  - 7\Delta/2} d ( e^{3\phi/8  + 7\Delta/2} s_\pm ) 
= \frac{\mu e^{\Phi_4}}{2} V_\pm^4  + 5 \Lambda K' ~ .
\end{equation}
The results presented in the following subsections can be derived in multiple ways, 
using various combinations of algebraic or/and differential conditions on spinor bilinears.

\subsection{Vanishing 4-form flux}

In this subsection we consider the case in which the 4-form flux $G$ is zero i.e. $\mu = 0$. 
We deduce that there are no supersymmetric AdS$_6$ solutions in this case. 
As mentioned earlier, if $\mu = 0$   also $s_\pm = 0$ and $K' =0$, the latter following immediately from \eqref{ders1}. Using this information, the algebraic conditions \eqref{algebraic} give additional constraints on scalar bilinears:
\begin{equation}
\Real(\eta^\dagger_1 \xi_1) = 0 \, , \quad
\eta^\dagger_1 \g_5 \xi_2 = 0 \, ,\quad 
\Imag(\eta^\dagger_1\g_5\xi_1) = 0 ~ .
\end{equation}
These constraints restrict the coefficients of the expansion of 
$\xi_i$ in terms of $\eta_i$ \eqref{expansion} in the following way
\begin{equation}\label{expan}
\xi_i = q \, \g_5 \eta_i + \textstyle{\sum_j} \, q_A \ii \sigma^A_{ij} \, \eta_j
\end{equation} 
where $q,q_A$ are real functions.

In an attempt to determine the geometry of $M_4$ we examine the differential conditions 
obeyed by the 1-form bilinears. We find that $\nabla_{(m} K_{n)} ^A =0$ and so we conclude 
that the dual vectors of $K^A$ are Killing vectors. Furthermore, we derive the differential equations
\begin{subequations}
\begin{align}
e^{2\Phi_4} d (e^{-2\Phi_4} K^A) &=   8\Lambda ( \mathrm{J}_+^A + \mathrm{J}_-^A ) \label{derK} \\                                                                                                     
e^{2\Phi_4} d (e^{-2\Phi_4} K)   &= \Real (\eta^\dagger_1 \g_5 \xi_1) \, m e^{\Phi_2} *_4 B' \label{KB'1} \\
dK &= 0 \label{KB'2}   
\end{align}
\end{subequations}
and $\mathscr{L}_{K_\sharp^A} \phi = \mathscr{L}_{K_\sharp^A} \D = 0$ where $\mathscr{L}$ denotes 
the Lie derivative and $\sharp$ the dual vector $g^{-1}(K^A, \ \cdot \ )$.
A natural question that arises is to determine the algebra of these Killing vectors. 
The Killing property of the vectors and the fact that their Lie derivative leaves invariant the dilaton $\phi$ 
and the warp factor $\D$ can be exploited and compute, using \eqref{derK} and Fierz identities, 
the commutator of the vectors. We derive
\begin{equation}
[K_\sharp^A, K_\sharp^B] = - 2 \Lambda (\hat{s}_+ + \hat{s}_-) \e^{ABC} K_\sharp^C
\end{equation}
The above equation leads to the conclusion that $(\hat{s}_+ + \hat{s}_-)$ is constant. 
A combination of this fact with \eqref{ders0} yields
\footnote{
We also require $\Hv \neq 0$; if $\Hv = 0$ then from the equation of motion for $H$ 
it follows that also $B'$ is zero and in that case a simple analysis of the 
supersymmetry conditions leads to $\Lambda = 0$.
} 
$\hat{s}_+ = \hat{s}_-$, $d\hat{s}_\pm = 0$ and 
\begin{subequations}
\begin{gather}
e^{\Phi_3} \Hv = -8 \Lambda K                                \\
m e^{\Phi_1} (V_+^4 - V_-^4) = -2\Lambda K  \label{KVrel}    \\
- \frac{1}{4} d\phi + 3 d\D = m e^{\Phi_1} (V_+^4 + V_-^4) \label{dilwarp1} 
\end{gather}
\end{subequations}
where since $\hat{s}_\pm$ are constant and equal we have set, without loss of generality, $\hat{s}_\pm = 1$. 
In order to further investigate these relations, we expand $K$ and $V_-^4$ in terms 
of the orthogonal 1-forms $V_+^a$, using the expansion \eqref{expan} of $\xi_i$ in terms of $\eta_i$. 
We derive
\begin{subequations}
\begin{align}
& \tfrac{1}{2} K =   q_A V_+^A - q V_+^4 \label{Kexp} \\
& V_-^4 = 2 q (q_A V_+^A - q V_+^4) + ( q^2  + q_A q_A) V_+^4 ~ .
\end{align}
\end{subequations}
Taking into account that $\hat{s}_- = (q^2 + q_A q_A) \hat{s}_+$ 
\begin{equation}
V_-^4 = q K + V_+^4 ~ . 
\end{equation}
Comparison with \eqref{KVrel} then yields $m q = 2 \Lambda e^{-\Phi_1}$. 
From the expansion \eqref{expan}, $q = \Real(\eta^\dagger_1\g_5 \xi_1)$ and for this scalar bilinear 
the supersymmetry conditions furnish
\begin{equation}
e^{\phi/4 - 3\D} d ( e^{-\phi/4 + 3\D} q ) = 4 \Lambda ( V_+^4 + V_-^4 ) ~ . 
\end{equation} 
Substituting the value $m q = \Lambda e^{-\Phi_1}$ derived above it follows that
\begin{equation}
-\frac{3}{2} d\phi + 2 d \D = 2 m e^{\Phi_1} ( V_+^4 + V_-^4 ) ~ .
\end{equation}
Combining this equation with \eqref{dilwarp1} yields $d\phi = -4 d \D$ and hence 
$\phi =  - 4 \D + c$ where $c$ is a constant.
For this value of the dilaton, $e^{-c/2} \Hv = - 8\Lambda K$ and so \eqref{KB'1} and 
the equation of motion \eqref{eom} become respectively
\begin{subequations}
\begin{align}
d(e^{8\D} \Hv) &= -16 e^{8\Delta} \Lambda^2 *_4 B' \\
d(e^{8\D} \Hv) &= -4 e^{3c/2} m^2 *_4 B'
\end{align}
\end{subequations}
We thus conclude that either $\Delta$ is constant or $B' = 0$. In the former case equations \eqref{KB'1} and \eqref{KB'2} lead to
\begin{equation}
 2\Lambda \, e^{-c/2} *_4 B'  = 0 ~ .
\end{equation}
Hence $B' =0$ in both cases. Since $H = dB'$, we also have $\Hv = K = 0$. 
From the expansion of $K$  \eqref{Kexp} in an orthogonal basis, we see that $q = q_A = 0$ and hence
$\xi_i = 0$, which is inconsistent with $\Lambda \neq 0$.

\subsection{Non-vanishing 4-form flux}

We start by showing that $s_-=s_+$. 
One way to derive this is as follows: from the algebraic conditions \eqref{alg3} and \eqref{alg4} 
one obtains $\Hv_m K'^m = 0$ 
while \eqref{alg1} and \eqref{alg2} yield $e^{\Phi_3} \Hv_m K'^m  = 6\Lambda(s_- - s_+)$. 
Hence $s_- = s_+$
\footnote{
Recall that $\mu\neq 0$ implies $s_\pm \neq 0$.
}.
Inspection of equations \eqref{key}, \eqref{ders0} and \eqref{ders1} then gives  
\begin{equation}\label{equalities}
\hat{s}_- = \hat{s}_+ \equiv \hat{s} \, , ~~~~
V_+^4 = V_-^4 \, , ~~~~
K = \Hv = 0 ~.
\end{equation}
For $\Hv = 0$,  the supersymmetry conditions \eqref{alg3} and \eqref{alg4} yield the equations
\begin{equation}
\pm d \phi \wedge V_\pm^4  = \frac{3me^{\Phi_2}}{2} \left( \hat{s}_\pm *_4 B'  -  s_\pm B' \right) ~ .
\end{equation}
Taking into account \eqref{equalities} and $s_-=s_+ \equiv s$ we arrive at  
\begin{equation}
B'(\hat{s}^2 - s^2) = 0. 
\end{equation}
Consequently, either $B'= 0$ or $\hat{s} = \pm s$. The latter case is equivalent to 
$\g_5 \eta_i = \pm \eta_i$ and $\g_5 \xi_i = \pm \xi_i$ (chiral spinors) and in this case,
a straightforward combination of the supersymmetry conditions 
\eqref{algebraic} and \eqref{differential} result in $\Lambda = 0$. Therefore, we proceed with the first case.

For $B' = 0$ and  $\Hv = 0$, from equations \eqref{alg1} or \eqref{alg2} and 
the symmetry properties of the 1-form and scalar bilinears, it follows that
$\eta^\dagger_1 \g_5 \xi_2 =  0 = \Imag(\eta^\dagger_1\g_5\xi_1)$. 
Then from equations \eqref{alg3} and \eqref{alg4} we derive
$\eta^\dagger_1 \xi_2 = 0 = \Imag(\eta^\dagger_1\xi_1)$. 
These constraints on the scalar bilinears, together with $s_- =s_+$ and $\hat{s}_- = \hat{s}_+$, 
restrict the coefficients of the expansion of $\xi_i$ in terms of $\eta_i$ \eqref{expansion} as 
\begin{equation}
a_2 = b_2 = 0 \, , \qquad a_1 = \pm 1 \, , \qquad b_1 = \pm 1 ~ .
\end{equation}
Equivalently $\xi_i = \pm \eta_i$ or $\xi_i = \pm \g_5 \eta_i$.  
Assuming $\eta_i = \pm \g_5 \xi_i$, the supersymmetry conditions give  $\Lambda = 0$. 
Then the only possibility left is $\xi_i = \pm \eta_i$
\footnote{
If an alternative decomposition of the $\mathrm{Cliff(9,1)}$ generators is chosen (c.f.\ appendix \ref{cliffspin}), 
then compatible with $\Lambda \neq 0$ is the choice $\xi_i = \pm \g_5 \eta_i$.
}. 
Different signs produce the same set of conditions with a different sign for $\Lambda$ and so, 
without loss of generality, one can choose either.

\section{The AdS$_6 \times_\w S^4$ solution}

\label{solution}

In this section we study the remaining case which 
- without loss of generality -  is $\eta_i = \xi_i$
\footnote{
Henceforth we omit the $\pm$ subscripts which are used to distinguish between 
bilinears of $\eta_i$ and bilinears of $\xi_i$.
}. 
It is convenient to write down the corresponding simplified set of supersymmetry
conditions. The differential conditions become 
\begin{equation}\label{diff}
\nabla_m \eta_i = \left[ \frac{\mu e^{\Phi_4}}{4} + \Lambda \right] \g_m \g_5 \eta_i   \, , \\
\end{equation}
 while  a set of reduced algebraic conditions is
\begin{subequations}
\begin{align}
& 0 = \partial_m \Delta \g^m \eta_i - \frac{m e^{\Phi_1}}{8} \eta_i + \frac{3\mu e^{\Phi_4}}{16} \g_5 \eta_i + 2 \Lambda \g_5 \eta_i  \label{als41} \\
& 0 = \partial_m \phi \, \g^m \eta_i +\frac{5m e^{\Phi_1}}{2}\eta_i + \frac{\mu e^{\Phi_4}}{4} \g_5 \eta_i ~ . \label{als42} 
\end{align}
\end{subequations}

From \eqref{diff} we find
\begin{equation}
d \hat{s} = 0 \, , \qquad
d s = - \left(\mu e^{\Phi_4} +4 \Lambda \right) V^4 
\end{equation}
and thus we can set $\hat{s} = 1$. Since $\hat{s} = \norm{\eta^+_i}^2 + \norm{\eta^-_i}^2$  
we introduce the parametrization 
\begin{equation}
\norm{\eta^+_i}^2 = \cos^2(\theta/2) \, ,
\qquad
\norm{\eta^-_i}^2 = \sin^2(\theta/2) \, ,
\qquad \theta \in (0, \pi/2) ~ .
\end{equation}
It follows that $s = \norm{\eta^+_i}^2 -  \norm{\eta^-_i}^2 = \cos\theta$. 
From the algebraic conditions \eqref{als41} and \eqref{als42} we then obtain 
\begin{subequations}
\begin{align}
- 2 m \, e^{\Phi_1} \cos\theta +3\mu \, e^{\Phi_4} + 32 \Lambda  &= 0 \\
10 m \, e^{\Phi_1}\cos\theta + \mu \, e^{\Phi_4} &= 0 ~ .
\end{align}
\end{subequations}
These relations lead to $\mu \, e^{\Phi_4} = -10 \Lambda$ and $m \, e^{\Phi_1} \cos\theta = \Lambda$. 
Therefore 
\begin{equation}
\D = - \frac{\phi}{20} + c
\end{equation}
where $c$ is a constant. Since $\Phi_1 = 5 \phi/4 + \D = 6 \phi/5$  we deduce
\begin{equation}
e^\phi = \left ( \frac{e^c m}{\Lambda} \cos\theta \right)^{-5/6} ~ .  
\end{equation}
With the above values, the  conditions \eqref{als41} and \eqref{als42} become identical. In addition we derive $\mathscr{L}_{V_\sharp^A} \phi =  0$.
The differential condition simplifies further and becomes
\begin{equation}\label{kill}
\nabla_m \eta_i = - \frac{3\Lambda}{2} \g_m \g_5 \eta_i  ~ .
\end{equation}
This implies that $M_4$ is an Einstein manifold with Ricci tensor  $\mathrm{Ric} = 3 \cdot (3\Lambda)^2 g$.
We recognise that (\ref{kill}) is the standard Killing spinor equation admitting  an $S^4$ as solution (c.f.\ \cite{Lu:1998nu}). Below we show that indeed the local Einstein metric on the round $S^4$ is the \emph{unique} solution to this equation. 
For the 1-form bilinears, equation \eqref{kill} yields 
\begin{equation}
\nabla_m \,V^A_n =  \frac{3 \Lambda}{2} \mathrm{J}^A_{mn} \, , \qquad
\nabla_m \,V^4_n  = - \frac{3\Lambda}{2} \, g_{mn} \cos{\theta} ~ .
\end{equation}
In particular,  the vector-duals of $V^A$ are Killing vectors whereas the vector-dual of $V^4$ is a conformal Killing vector.  Taking into account the expressions of $\mathrm{J}^A$ \eqref{JV} in terms of $V^a$ and the differential equation $d\cos \theta = 6\Lambda V^4$ obtained earlier, we derive 
\begin{equation}
d V^A = \frac{12\Lambda}{\sin^2 \theta} \frac{1}{2} \e^{ABC} V^B \wedge V^C 
      - \frac{1}{\sin^2 \theta} \, V^A \wedge d ( \sin^2 \theta ) 
\end{equation}
Setting  
\begin{equation}
\hat{\sigma}^A \equiv - \frac{12\Lambda}{\sin^2\theta} V^A \, , \qquad 
\end{equation}
the above equation becomes
\begin{equation}
d\hat{\sigma}^A = - \frac{1}{2} \e^{ABC} \hat{\sigma}^B \wedge \hat{\sigma}^C ~ .
\end{equation}
We can thus identify $\hat{\sigma}^A$ as the left-invariant forms on $S^3$
\begin{equation}\label{coor}
\hat{\sigma}_1 + i \hat{\sigma}_2 = e^{-i\psi}(d\vartheta + i\sin\vartheta d\phi) ~,~~~ \hat{\sigma}_3 = d\psi + \cos\vartheta d\phi ~.
\end{equation}

The dual Killing vectors $V^A_\sharp$ which obey the $\mathfrak{su}(2)$ algebra\footnote{More accurately it is $-\frac{1}{3\Lambda}V^A_\sharp$ that obey the canonical $\mathfrak{su}(2)$ commutation relations and generate the right action of $SU(2)$.}
\begin{equation}
[V^A_\sharp, V^B_\sharp] = - 3\Lambda \, \e^{ABC} V^C_\sharp
\end{equation}
are identified as the generators of $\mathrm{SU(2)_R}$. In particular, the Killing spinor $\eta_i$ transforms under $\mathfrak{su}(2)_{\mathrm{R}}$ as
\begin{equation}
\mathscr{L}_{V_\sharp^A} \eta_i = \frac{3\Lambda}{2} \, \textstyle{\sum}_j \ii \sigma_{ij}^A \, \eta_j ~ ,
\end{equation}
where $\mathscr{L}_{V_\sharp^A}$ is the spinorial Lie derivative
\begin{equation}
\mathscr{L}_{V_\sharp^A} = V_\sharp^{Am} \nabla_m + \tfrac{1}{4} \nabla^{[m} V_\sharp^{A n]} \g_{mn} ~.
\end{equation}

The metric on $M_4$ constructed out of the orthonormal frame 
\begin{equation}\label{orframe}
e^4 = \frac{1}{3\Lambda} d\theta~,~~~~~~~ e^A = \frac{1}{6\Lambda} \sin\theta \hat{\sigma}^A
\end{equation}
defined by $V^a$  takes the form
\begin{equation}
ds_4^2 =  
\frac{1}{(3\Lambda)^2} \left[ d\theta^2 +  \frac{1}{4} \displaystyle \sin^2\theta  \sum_A (\hat{\sigma}^A)^2  \right] 
\equiv \frac{1}{(3\Lambda)^2} d\Omega^2_4 ~ ,
\end{equation}
where $\tfrac{1}{4}\sum_A (\hat{\sigma}^A)^2$ is the round metric on $S^3$. 
The complete 10-dimensional solution reads
\begin{subequations}\label{10solution}
\begin{gather}
e^\phi = \left( \frac{e^c m}{\Lambda} \cos\theta \right)^{-5/6} \\
G = - \frac{5}{12} \frac{e^{3c}}{(3\Lambda)^3} \, e^{-2\phi/5} \, \sin^3\theta \, d\theta \wedge \mathrm{vol}_{S^3} \\
ds_{10}^2 = e^{-\phi/10} \frac{e^{2c}}{(3\Lambda)^2} 
\left \{ \frac{9}{4} ds^2_{\mathrm{AdS}_6}  + d\Omega^2_4 \right \} ~ ,
\end{gather}
\end{subequations}
where $\mathrm{vol}_{S^3}$ is the volume element of the unit 3-sphere and $ds^2_{\mathrm{AdS}_6}$ is the line element of unit AdS$_6$.
The AdS$_6 \times_\w S^4$ solution as presented here has the same form as in \cite{Pope} upon 
\begin{equation}
\theta \rightarrow \pi/2 -\theta \, , ~~~~
\Lambda = e^c m \, , ~~~~
m \rightarrow m/2 
\end{equation} 
and as in \cite{Oz} upon transforming the metric to the string frame $g_{string} = e^{\phi/2} g_{Einstein}$ and 
\begin{equation}
\theta \rightarrow \pi/2 - \alpha \, , ~~~~
m \rightarrow m/2 \, , ~~~~
\frac{e^c}{3\Lambda} \equiv Q^{3/10}_4 C^{-1/5} ~ ,
\end{equation}
where we have set the string length $l_s = 1$.

As discussed in \cite{Oz}, this solution has a boundary at $\theta= \pi/2$, corresponding to the equator of $S^4$. 
Hence $M_4$ is a hemisphere, the boundary equator of which was identified in \cite{Oz} with an orientifold plane. 
We note that at the (north) pole, corresponding to $\theta = 0$, one chiral component of the spinor $\eta_i$ vanishes,
while on the equator, corresponding to $\theta =\pi/2$, the  chiral components of $\eta_i$ have equal norms. 

In the frame \eqref{orframe} and upon substituting the value of the dilaton, the algebraic condition \eqref{als42} becomes 
\begin{equation}\label{projection}
\textstyle{\frac{1}{2}} (1 + \sin\theta \g_4 - \cos\theta \g_5) \eta_i = 0 ~ .
\end{equation}
The operator acting on $\eta_i$ has the properties of a projection operator, 
reducing the independent components of $\eta_i$ by half. In particular, in the representation
\begin{equation}\label{Cliffrep}
\g_4 = \begin{pmatrix} 0 & \mathbb{I}_2 \\ \mathbb{I}_2 & 0 \end{pmatrix} ~,~~~~~
\g_A = \begin{pmatrix} 0 & -\ii \sigma^A \\ \ii \sigma^A & 0 \end{pmatrix} 
\end{equation}
of the generators of $\mathrm{Cliff(4,0)}$, the condition \eqref{projection} becomes
\begin{equation}\label{spinrel}
\eta^-_i \cos(\theta/2) = - \eta^+_i \sin(\theta/2) ~.
\end{equation}

Taking into account \eqref{spinrel}, and in the frame \eqref{orframe}, one can solve the Killiing spinor equations \eqref{kill} and 
recover all the Killing spinors. We find
\begin{equation}\label{Ksint}
\eta^+_1 = \cos(\theta/2) \begin{pmatrix} \ell_1 \\ \ell_2 \end{pmatrix} ~,~~~~
\eta^-_1 = - \sin(\theta/2) \begin{pmatrix} \ell_1 \\ \ell_2 \end{pmatrix}
\end{equation}
where $\ell_1$ and $\ell_2$ are complex constants. 
Accordingly, the components of the ten-dimensional Killing spinor
\begin{equation}
\e = \sum_{i=1}^2 \psi_i \otimes \eta_i
\end{equation}
(where $\psi_i$ are the Killing spinors on AdS$_6$ \cite{Lu:1998nu}) are reduced by half i.e.\ $\e$ has 16 real independent components, 
in accordance with the number of supercharges of the ${F(4)}$ superalgebra. Moreover, since the spinors $\eta_i$ transform in the
 $(\mathbf{2,1})$ representation of the ${SO(4)} \simeq {SU(2)_R} \times {SU(2)}$ isometry subgroup, one can consider 
orbifolds $S^4/\Gamma$ where $\Gamma$ is an ADE subgroup of ${SU(2)} \subset {SO(4)}$, without furher breaking supersymmetry. 
 An explicit example is a $\mathbb{Z}_n$ quotient acting on the coordinate $\psi$ introduced in \eqref{coor}, which leaves supersymmetry intact since 
the Killing spinors \eqref{Ksint} do not depend on $\psi$. It is interesting to note that in this case one might consider turning on a 
RR two-form flux $F = dA$ through the vanishing $S^2$ proportional to $m B$, so that $B'=0$ - see equation \eqref{B'}. This is a ``flat'' deformation of the geometry, 
that does not alter the form of the metric\footnote{I would like to thank Diego Rodriguez-Gomez for pointing this out.}.

\section{Conclusions}

In this note we have performed a systematic analysis of general supersymmetric AdS$_6$ backgrounds of massive type IIA supergravity. We have established the uniqueness of the 
AdS$_6 \times_{\mathrm{w}} S^4$  solution of \cite{Oz}, and  certain orbifolds, and discussed its supersymmetric properties. 
Although the present work does not exclude 
the existence of supersymmetric AdS$_6$ vacua in other supergravity theories, it suggests  a scarcity of such backgrounds.
In particular, we do not expect that supersymmetric AdS$_6$ solutions can be found in type IIA or eleven dimensional supergravity.

\subsection*{Acknowledgements}
I would like to thank Dario Martelli for suggesting this research topic and for guidance through all the stages of the present work, and James Sparks and Diego Rodriguez-Gomez for useful comments. My work is supported by an A.G. Leventis Foundation grant, an STFC studentship and via
the Act ``Scholarship Programme of S.S.F. by the procedure of individual
assessment, of 2011-12'' by resources of the Operational Programme for Education
and Lifelong Learning, of the European Social Fund (ESF) and of the NSRF,
2007-2013.

\appendix

\section{Spinors and Clifford algebras} 

\label{cliffspin}

\subsubsection*{Clifford algebra and spinors in 4 dimensions}

The (complex) Clifford algebra in four Euclidean dimensions $\mathrm{Cliff(4,0)}$ is isomorphic to the matrix algebra of $4 \times 4$ complex matrices $\mathrm{Mat}_4(\mathbb{C})$. The generators of $\mathrm{Cliff(4,0)}$ satisfy $\{\gamma_a, \gamma_b \} = 2 \delta_{ab}$ and are chosen so that  $\gamma_a^\dagger = \gamma_a$. The chirality operator is $\g_5 = \g_1 \g_2 \g_3 \g_4$ and has the property $\g^2_5 = \mathbb{I}$.
The intertwiners that relate the representations $\{ \g_a, \ \g^T_a, \ \g^*_a \}$ of $\mathrm{Cliff(4,0)}$ are
\begin{equation}
C_4 \, \gamma_a \, C^{-1}_4 =  \gamma_a^T \, , \qquad
B_4 \, \gamma_a \, B^{-1}_4 =  \gamma_a^* ~ .
\end{equation}
They satisfy $C_4 = - C^T_4$,  $B^*_4 B_4 = -\mathbb{I}$  and  are related as $C_4 = B^T_4$.

Under $\mathrm{Spin(4)} \subset \mathrm{Cliff(4,0)}$ an irreducible representation of $\mathrm{Cliff(4,0)}$ (Dirac spinor) decomposes into two irreducible $\mathrm{Spin(4)}$ representations of opposite chirality (Weyl spinors). The charge-conjugate $\eta^c$ of a spinor $\eta$ is defined as
$\eta^c \equiv B_4^{-1} \eta^*$ and obeys the relation $\eta^{cc} = -\eta$. Setting $\eta_1 \equiv \eta$ and $\eta_2 \equiv \eta^c$, the aforementioned relation can be summarised as
\begin{equation}
\eta^c_i = \textstyle{\sum}_j \epsilon_{ij} \,\eta_j 
\end{equation}
where $\e_{ij}$ is antisymmetric in $i,j$ and $\e_{12} = 1$. This symplectic-Majorana property is compatible with the chirality condition.

\subsubsection*{Clifford algebra and spinors in 5+1 dimensions}
The (complex) Clifford algebra in $5+1$ dimensions $\mathrm{Cliff(5,1)}$ is isomorphic to the matrix algebra of $8 \times 8$ complex matrices $\mathrm{Mat}_8(\mathbb{C})$. The generators of $\mathrm{Cliff(5,1)}$ satisfy $\{\g_\alpha, \g_\beta \} = 2 \eta_{\alpha\beta}$ and are chosen so that
$\g_\alpha^\dagger = \g_0\,\g_\alpha\,\g_0$. The chirality operator is  $\g_7 = \g_0 \g_1 \g_2 \g_3 \g_4 \g_5$ and has the property $\g^2_7 = \mathbb{I}$.
The intertwiners that relate the representations $\{\g_\alpha, \ -\g^T_\alpha, \ \g^*_\alpha \}$ of $\mathrm{Cliff(5,1)}$ are
\begin{equation}
C_6 \, \g_\alpha \, C^{-1}_6 = -\g_\alpha^T \, , \qquad
B_6 \, \g_\alpha \, B^{-1}_6 =  \g_\alpha^*  ~ .
\end{equation}
They have the properties $C_6 = C^T_6 $, $ B^*_6 B_6 = -\mathbb{I}$ and are related as $C_6 = B^T_6 \g_0$.

Under $\mathrm{Spin(5,1)} \subset \mathrm{Cliff(5,1)}$ an irreducible representation of $\mathrm{Cliff(5,1)}$ (Dirac spinor) decomposes to two irreducible $\mathrm{Spin(5,1)}$ representations of opposite chirality (Weyl spinors). The charge-conjugate $\psi^c$ of a spinor $\psi$ is defined as
$\psi^c \equiv B_6^{-1} \psi^*$ and obeys the relation $\psi^{cc} = -\psi$. Setting $\psi_1 \equiv \psi$ and $\psi_2 \equiv \psi^c$, the aforementioned relation can be summarised as
\begin{equation}
\psi^c_i = \textstyle{\sum}_j \epsilon_{ij} \, \psi_j ~.
\end{equation}
This symplectic-Majorana property is compatible with the chirality condition.

\subsubsection*{Clifford algebra and spinors in 9+1 dimensions}

The (complex) Clifford algebra in $9+1$ dimensions $\mathrm{Cliff(9,1)}$ is isomorphic to the matrix algebra of $32 \times 32$ complex matrices $\mathrm{Mat}_{32}(\mathbb{C})$. The generators of $\mathrm{Cliff(9,1)}$ satisfy $\{\Gamma_A, \Gamma_B \} = 2 \eta_{AB}$ and are chosen so that $\Gamma_A^\dagger = \Gamma_0\,\Gamma_A\,\Gamma_0$. The chirality operator is $\Gamma_{11} = \Gamma_0 \Gamma_1\dots \Gamma_9$ and has the property $\Gamma^2_{11} = \mathbb{I}$. 
The intertwiners that relate the representations $\{ \G_A, \ -\G^T_A, \ \G^*_A \}$ of $\mathrm{Cliff(9,1)}$ are 
\begin{equation}
C_{10} \, \Gamma_A \, C^{-1}_{10} = -\Gamma_A^T \, ,\qquad
B_{10} \, \Gamma_A \, B^{-1}_{10} =  \Gamma_A^*  ~ .
\end{equation}
They have the properties $C_{10} = - C^T_{10}$, $B^*_{10} B_{10} = \mathbb{I}$ and are related as $C_{10} = B^T_{10} \Gamma_0$.

Under $\mathrm{Spin(9,1)} \subset \mathrm{Cliff(9,1)}$ an irreducible representation of $\mathrm{Cliff(9,1)}$ (Dirac spinor) decomposes to two irreducible $\mathrm{Spin(9,1)}$ representations of opposite chirality (Weyl spinors). Consistent with the chirality condition is the Majorana property 
$\epsilon^* = B_{10}\, \epsilon$.

\subsubsection*{$\mathrm{Cliff}(9,1) \simeq \mathrm{Cliff(5,1)} \otimes \mathrm{Cliff(4,0)}$ decomposition}
The generators of $\mathrm{Cliff}(9,1)$ are decomposed as
\begin{equation}
 \Gamma_\alpha = \g_\alpha \otimes \gamma_5 \quad \text{and} \quad 
 \Gamma_{a+5} = \mathbb{I} \otimes \gamma_a  ~ .
\end{equation}
where $\alpha \in \{0,\dots, 5\}$ and $a \in \{ 1,2,3,4 \}$ are tangent space indices. Accordingly, the decomposition of the intertwiners is $ C_{10} = C_6 \otimes C_4$ and  $B_{10} = B_6 \otimes B_4$ and of the chirality operator $\Gamma_{11} = \g_7 \otimes \gamma_5$.
There is also an alternative decomposition $\G_\alpha = \g_\alpha \otimes \mathbb{I}$ and $\G_{a+5} = \g_7 \otimes \gamma_a$  which is related to the above via a similarity transformation 
$U \equiv P_-\otimes \gamma_5 + P_+\otimes \mathbb{I}$  where $P_{\pm} = \frac{1}{2} (\mathbb{I} \pm \g_7)$.

\section{Spinor bilinears of $\mathrm{Cliff(4,0)}$} \label{bilinearsproperties}

The algebra $\mathrm{Cliff(4,0)}$ is spanned by the elements
$\{ \mathbb{I}, \ \g_a, \ \g_{ab}, \ \g_{abc}, \ \g_{abcd} \}$, which are subject to the relations
\begin{equation}
\g_{abc} = - \e_{abcd} \g^d \g_5 \, ,\qquad 
\g_a = \frac{1}{3!} \e_{abcd} \g^{bcd} \g_5  \, ,\qquad 
\g_{ab} = -\frac{1}{2!} \e_{abcd} \g^{cd} \g_5 ~ .
\end{equation}
The relation $C_4 = B^T_4$ leads to 
\begin{equation}
\eta_i^\dagger = \sum_{j=1}^2 \e_{ij} \, \eta^T_j C_4
\end{equation}
Bilinears of the form $\psi^T_i C_4 \gamma_{(n)} \chi_k$ obey the reality conditions
\begin{equation}\label{reality}
(\psi^T_1 C_4 \gamma_{(n)} \chi_2)^* = -\psi^T_2 C_4 \gamma_{(n)} \chi_1 \, , \qquad 
(\psi^T_1 C_4 \gamma_{(n)} \chi_1)^* =  \psi^T_2 C_4 \gamma_{(n)} \chi_2 ~ .
\end{equation}
Furthermore, the transposition identity
\begin{equation}
(C_4 \g^{a_1 \dots a_n})^T = -(-1)^{n(n-1)/2} C_4 \g^{a_1 \dots a_n} \, ,
\end{equation}
which follows from the properties of $C_4$, yields
\begin{equation}
\psi^T_i C_4 \gamma_{(n)} \chi_k = (\psi^T_i C_4 \gamma_{(n)} \chi_k)^T 
                                                      = -(-1)^{n(n-1)/2} \chi^T_k C_4 \gamma_{(n)} \psi_i  ~ .
\end{equation}
The Fierz identity for $\mathrm{Cliff(4,0)}$ reads
\begin{equation}
\begin{split}
\chi \psi^\dagger 
= \frac{1}{4} ( \psi^\dagger \chi + \g_a \,\psi^\dagger \g^a \chi - \frac{1}{2!} \g_{ab} \, \psi^\dagger \g^{ab}\chi 
               -\g_a \g_5 \, \psi^\dagger \g^a \g_5 \chi +\g_5 \, \psi^\dagger \g_5 \chi ) ~ .
\end{split}
\end{equation}


\begin{thebibliography}{1}


\bibitem{Maldacena}
J. Maldacena, 
``The large $N$ limit of superconformal field theories and supergravity'',
Adv. Theor. Math. Phys. 2 (1998) 231, hep-th/9711200

\bibitem{Martelli:2003ki}
  D. Martelli and J. Sparks,
  ``$G$-structures, fluxes and calibrations in M theory,''
  Phys. Rev. D 68, 085014 (2003), hep-th/0306225

\bibitem{AdS5_M}
J. P. Gauntlett, D. Martelli, J. Sparks and D. Waldram, 
``Supersymmetric AdS$_5$ solutions of M-theory'', 
Class. Quant. Grav. 21 (2004) 4335, hep-th/0402153

\bibitem{Lin:2004nb} 
  H. Lin, O. Lunin and J. M. Maldacena,
  ``Bubbling AdS space and 1/2 BPS geometries','
  JHEP 0410, 025 (2004), hep-th/0409174

\bibitem{Tsimpis}
D. Lust and D. Tsimpis, 
``Supersymmetric AdS$_4$ compactifications of IIA supergravity'',
JHEP02 (2005) 027, hep-th/0412250

\bibitem{AdS5_IIB}
J. P. Gauntlett, D. Martelli, J. Sparks and D. Waldram,
``Supersymmetric AdS$_5$ solutions of type IIB supergravity'',
Class. Quantum Grav. 23 (2006) 4693, hep-th/0510125

\bibitem{AdS4_M}
M. Gabella, D. Martelli, A. Passias, J. Sparks,	
``${\cal N}=2$ supersymmetric AdS$_4$ solutions of M-theory'',
hep-th/1207.3082

\bibitem{Seiberg:1996bd}
N. Seiberg,
``Five-dimensional SUSY field theories, nontrivial fixed points and string dynamics'',
Phys. Lett. B 388 (1996) 753, hep-th/9608111

\bibitem{Morrison:1996xf}
D. R. Morrison and N. Seiberg,
``Extremal transitions and five-dimensional supersymmetric field theories'',
Nucl. Phys.  B  483 (1997) 229, hep-th/9609070

\bibitem{Rodriguez-Gomez}
O. Bergman, D. Rodriguez-Gomez,
``5d quivers and their AdS$_6$ duals'',
hep-th/1206.3503

\bibitem{KKL}
Hee-Cheol Kim, Sung-Soo Kim, Kimyeong Lee,
``5-dim Superconformal Index with Enhanced $E_n$ Global Symmetry'',
hep-th/1206.6781

\bibitem{Jafferis}
D. L. Jafferis, S. S. Pufu,
``Exact results for five-dimensional superconformal field theories with gravity duals'',
hep-th/1207.4359

\bibitem{Oz}
A. Brandhuber and Y. Oz, 
``The D4-D8 Brane System and Five Dimensional Fixed Points'', 
Phys.Lett. B460 (1999) 307-312, hep-th/9905148

\bibitem{Romans}
L. J. Romans, 
``Massive N=2a Supergravity In Ten-Dimensions" 
Phys. Lett. B 169 (1986) 374

\bibitem{Ferrara}
S. Ferrara, A. Kehagias, H. Partouche and A. Zaffaroni, 
``AdS$_6$ interpretation of $5D$ superconformal field theories'', 
Phys. Lett. B431 (1998) 57, hep-th/9804006

\bibitem{RomansF(4)}
L.J. Romans, 
``The $F(4)$ gauged supergravity in six dimensions'', 
Nucl. Phys. B269 (1986) 691

\bibitem{Pope}
M. Cvetic, H. Lu and and C.N. Pope, 
``Gauged Six-dimensional Supergravity from Massive Type IIA'',
Phys. Rev. Lett. 83 (1999) 5226-5229, hep-th/9906221

\bibitem{D'Auria}
R. D'Auria, S. Ferrara and S. Vaula,
``Matter coupled $F(4)$ supergravity and the AdS$_6$/CFT$_5$ correspondence'',
JHEP10 (2000) 013, hep-th/0006107

\bibitem{Nahm}
W. Nahm, 
``Supersymmetries and their representations", 
Nucl. Phys. B 135 (1978) 149

\bibitem{FigueroaO'Farrill:1999va}
 J. M. Figueroa-O'Farrill,
 ``On the supersymmetries of Anti-de Sitter vacua'',
Class. Quant. Grav.  16  (1999) 2043, hep-th/9902066

\bibitem{FigueroaO'Farrill:2007ar}
J. M. Figueroa-O'Farrill, E. Hackett-Jones and G. Moutsopoulos,
``The Killing superalgebra of ten-dimensional supergravity backgrounds'',
Class. Quant. Grav. 24  (2007) 3291, hep-th/0703192

\bibitem{Lu:1998nu}
H. Lu, C. N. Pope and J. Rahmfeld,
``A Construction of Killing spinors on $S^n$'',
J. Math. Phys. 40  (1999) 4518, hep-th/9805151


\end{thebibliography}
\end{document}